\documentclass[letter]{aa} 

\usepackage{graphicx}
\usepackage{subcaption}
\usepackage[colorlinks=true,     linkcolor=blue, citecolor=blue, filecolor=blue, urlcolor=blue]{hyperref}
\usepackage{txfonts}
\begin{document}

	\title{Low- and high-velocity `water fountains': different evolutionary stages}

    \titlerunning{Low- and high-velocity `water fountains'}
	\authorrunning{R.~A. Cala et al.}
    
	\author{Rold\'an A. Cala \inst{1}
          \and          
         Luis F. Miranda \inst{1}
          \and
          Jos\'e F. G\'omez \inst{1}
          \and
          Keiichi Ohnaka \inst{2}
                    }
	
	\institute{Instituto de Astrof\'{\i}sica de Andaluc\'{\i}a, CSIC, 	Glorieta de la Astronom\'{\i}a s/n, E-18008 Granada, Spain\\
              \email{roldancalab@gmail.com}
         \and
         Instituto de Astrof\'{\i}sica, Departamento de F\'{\i}sica y Astronom\'{\i}a, Facultad de Ciencias Exactas, Universidad Andrés Bello, Fernández
         Concha 700, Las Condes, Santiago, Chile
             }
	
	
	\abstract{`Water fountains' (WFs) are optically obscured evolved stars, most of them thought to be in the post-asymptotic giant branch (post-AGB) phase, characterized by H$_{2}$O maser emission tracing molecular jets. Interestingly, four WFs (IRAS\,15445$-$5449, IRAS\,18019$-$2216, IRAS\,18443$-$0231, and IRAS\,18464$-$0140) and one WF candidate (IRAS\,18480+0008) are potential planetary nebulae (PNe) because they exhibit radio continuum emission, suggesting the presence of a photoionized region characteristic of PNe. To classify these objects, we obtained \textit{K}-band (2.0-2.3\,$\mu$m) spectra of these WFs, including the only WF PN known (IRAS\,15103$-$5754) for comparison. Our spectra reveal two group of sources: (i) `low-velocity' WFs with H$_2$O maser velocity spread of $\lesssim$50\,km\,s$^{-1}$ (IRAS\,18019$-$2216, IRAS\,18464$-$0140, and IRAS\,18480+0008) showing the CO band at 2.29\,$\mu$m in absorption, typical of cool giant stars, and no emission lines; and (ii) `high-velocity' WFs, velocity spread of $\gtrsim$50\,km\,s$^{-1}$ (IRAS\,15103$-$5754, IRAS\,15445$-$5449, and IRAS\,18443$-$0231), exhibiting emission lines of Br$\gamma$, He\,{\sc i}, and H$_2$, consistent with hotter central stars and/or shock-excited emission. The emission line ratios of these lines in IRAS\,18443$-$0231 indicates that it may be a nascent PN. The spectrum of IRAS\,15445$-$5449 also shows a CO band and Na\,{\sc i} doublet in emission, suggesting the presence of a compact circumstellar disk and/or active mass loss. These results favor the previously suggested notion that the difference between low- and high-velocity WFs is not simply a projection effect but reflects intrinsically different evolutionary stages. Moreover, the results are also consistent with the idea of an increase in the jet ejection velocity as the post-AGB evolution proceeds.}

	
	\maketitle
	%
	
	\section{Introduction}

    `Water fountains' (WFs) are evolved stars, most of them believed to be in the post-asymptotic giant branch (post-AGB) phase, with H$_{2}$O maser emission tracing collimated mass loss. High-resolution observations of these H$_2$O masers show incipient small-scale ($\sim$10$^{2}$-10$^{3}$\,au), bipolar jets that can reach velocities of hundreds of km\,s$^{-1}$ \citep[e.g.,][]{ima02, bob05, gom11, usc23}. The short kinematical ages estimated for WF jets \citep[$\la 100$ yr; e.g.,][]{day10, yung11, oro19, taf20} suggest that these jets could be one of the initial manifestations of non-spherical mass loss that occur in evolved low- and intermediate-mass stars \citep[e.g.,][]{ima02}. So far, 18 objects have been confirmed as WFs \citep[][]{gom17, cal25}. Among these, only one (W43A) has been classified as an AGB star \citep{ima02, ima05}, and one (IRAS\,15103$-$5754) as a PN, which could be one of the youngest PNe known \citep{gom15}.    

    Recently, four additional interesting objects were identified: three new confirmed WFs (IRAS\,18019$-$2216, IRAS\,18443$-$0231, and IRAS 18464$-$0140) and a new WF candidate, IRAS\,18480+0008 \citep{cal25}. Remarkably, these four objects present radio continuum emission indicating the presence of an ionized region that characterizes PNe. Thus, they could be new WF PNe. However, a definitive classification as PNe requires further spectroscopic confirmation, since shock-ionized gas around post-AGB stars can also produce radio continuum emission \citep[e.g.][]{cer17, ps17}. We note that, in addition to these four objects and the WF PN IRAS 15103$-$5754, only another WF IRAS\,15445$-$5449 presents radio continuum emission and its evolutionary stage is unknown \citep[][]{ps13}. Interestingly, the radio continuum emission in IRAS\,15445$-$5449, IRAS\,15103$-$5754, and IRAS\,18443$-$0231 is not mainly due to free-free processes, but it is of non-thermal nature \citep{ps13, sua15, cal25}. 

    Most WFs are optically obscured \citep[e.g.][]{sua08, gom17, ps17} because they are surrounded by thick circumstellar envelopes, probably as a result of a common-envelope episode in a close binary system \citep[][]{kho21}. Therefore, their characterization requires observations at infrared (or longer) wavelengths. 
        
    In this Letter, we present near-infrared spectra of the four radio continuum emitting WFs (IRAS 15445$-$5449, IRAS 18019$-$2216, IRAS 18443$-$0231, and IRAS 18464$-$0140) and the WF candidate (IRAS\,18480+0008) aiming at clarifying their nature. We also observed the WF PN IRAS\,15103$-$5754 for comparison purposes.

    \section{Observations and data processing}
    \label{sec:obs}
    
    We have performed long-slit intermediate resolution \textit{K}-band near-infrared spectroscopy with the New Technology Telescope (NTT) of the European Southern Observatory (ESO) located on La Silla Observatory in Chile on 11 and 12 June 2023 (Project ID: 111.24T1.001; P.I.: R.~A.\,Cala). Photometric conditions, average temperatures of +10$\degr$C and seeing of 2$\arcsec$ were experienced on both nights. To obtain and record the spectra, we used the infrared spectrograph Son of ISAAC (SOFI), a slit of 1$\arcsec$ width and 1.4$\arcmin$ length, the third order HR grism (resolution of R$\simeq$2200 for a 0.\arcsec6 slit width and wavelength range 2.00-2.30\,$\mu$m), and the 1024$\times$1024 Hawaii HgCdTe array. The position angle of the slit was 0$^{\degr}$ in all cases but in IRAS\,15103$-$5754, where the slit was oriented along the jet traced by the H$_{2}$O masers \citep[+56$^{\degr}$;][]{gom15}. In order to cancel out the telluric lines, we observed B main-sequence stars after or before each target, making use of the same instrumental setup (see Appendix\,\ref{ap:obs}). 

    The spectra of the telluric standard stars and of our targets were processed using the ESO's SOFI pipeline in \textit{esorex}, which performed dark-current correction, flat fielding, slit curvature distortion, combination of frames, and wavelength calibration. The dispersion obtained was $\sim$5$\AA$\,pixel$^{-1}$. The wavelength accuracy is $\sim$0.004\,$\mu$m in most of the spectral range, and $\sim$0.01\,$\mu$m longer than 2.25\,$\mu$m. Then, we used {\scriptsize IRAF} to extract the spectra and clean the bad pixels. We then removed the Br$\gamma$ absorption line from the telluric standards by interpolating across that part of the spectrum. We then divided the target spectra by those of their corresponding telluric standard, in order to cancel out atmospheric absorption features. The absolute flux calibration was carried out by multiplying by a normalized blackbody spectrum at the temperature of the telluric standard, and the result was re-scaled to fit the \textit{K}-band magnitudes of our targets reported in Two Micron All Sky Survey (2MASS) and United Kingdom Infrared Deep Sky Survey (UKIDSS). Assuming that the telluric standards are not variable in the infrared, uncertainties of $\pm$5-10$\%$ in absolute flux calibration are expected from this method, although taking into account that the slit width could be narrower than the angular size of the observed targets, some emission could be filtered out, thus larger uncertainties are possible.

\section{Results and discussion}
\label{sec:results}

\begin{figure*}
      \centering
      \includegraphics[width=0.99\hsize]{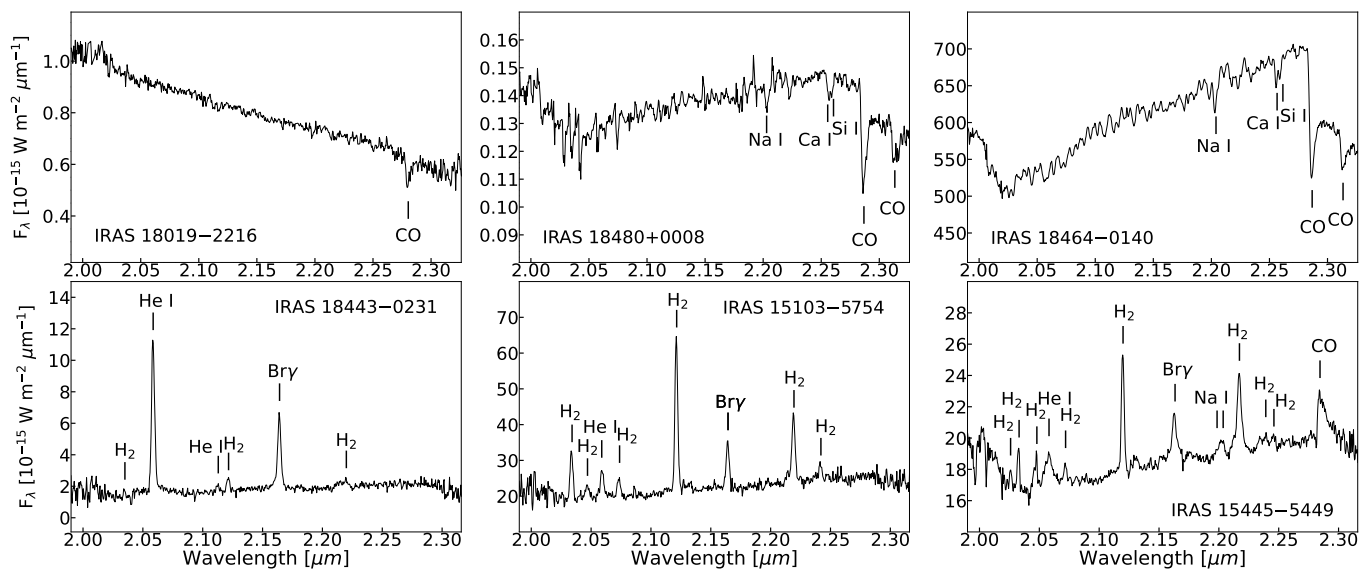}
      \caption{\textit{K}-band near-infrared spectra of our targets. The top panels are the objects without emission lines (Group A) and the bottom panels are the objects with emission lines (Group B). Other possible emission lines are are difficult to identify due to the low resolution of the spectra.}
      \label{fig:wf_spectra}
\end{figure*}

Fig.\,\ref{fig:wf_spectra} shows the spectra of the six observed WFs, which allow us to classify the objects in two groups. The sources in the top panels of Fig.\,\ref{fig:wf_spectra} (IRAS\,18019$-$2216, IRAS\,18464$-$0140, and IRAS\,18480+0008; hereafter Group A) show no emission lines but only absorption features due to the first overtone bandhead of CO ($\nu$\,=\,2--0) at 2.29\,$\mu$m and CO\,($\nu$\,=\,3--1) at 2.31\,$\mu$m, although the last is not detected in IRAS\,18019$-$2216. Weak absorption lines of Na\,{\sc i}, Ca\,{\sc i}, and Si\,{\sc i} at $\sim$2.20, $\sim$2.25, and $\sim$2.26 $\mu$m, respectively, are detected in IRAS\,18480+0008 and, apparently, also in IRAS\,18464$-$0140. 

In the sources shown in the bottom panels of Fig.\,\ref{fig:wf_spectra} (IRAS\,15103$-$5754, IRAS\,15445$-$5449, and IRAS\,18443$-$0231; hereafter Group B), the spectra are rich in emission lines and there is no obvious CO absorption. In IRAS\,15445$-$5449, there is emission of the CO ($\nu$\,=\,2--0) bandhead and of the doublet of Na\,I at 2.206 and 2.209\,$\mu$m observed around 2.210 $\mu$m. In Appendix\,\ref{ap:lines} and Table\,\ref{tab:lines} we report fluxes of the emission lines detected in Group B. We note that the spectrum of the WF IRAS\,18443$-$0231 (Fig.\,\ref{fig:wf_spectra}, bottom left panel) is consistent with the spectra of this source previously reported in near-infrared surveys \citep{coo13, kan15}.

\subsection{Evolutionary stage of the objects}
\label{sec:ev_stage}

The most noticeable spectral features in the spectra of sources in Group A are the absorption of the CO bands near 2.29 and 2.31 $\mu$m. These bands are characteristic of cool ($T_{\rm eff}$$\simeq$2100-5700\,K) main-sequence, giant, and supergiant stars of spectral type G, K, and M \citep[][]{ray09}. They are ubiquitous in
the spectra of AGB stars and have been detected in some early post-AGB stars \citep[e.g.,][]{hri94, oud95, kan15}. We also notice a rising continuum towards longer wavelengths in IRAS 18464$-$0140 and IRAS 18480+0008.

By comparing the spectra of our Group A sources with those in the atlas of infrared spectra by \citet[][]{ray09}, these objects should be M-type giants or supergiants. This is consistent with an AGB or early post-AGB nature. Thus, although
the radio continuum emission in the Group A sources indicates the presence of ionized gas, we do not expect this ionization to be due to ultraviolet radiation, as in the case of PNe, but to shocks, as previously reported in post-AGB stars \citep[e.g.][]{cer17, ps17}.

The spectra of sources in Group B are dominated by recombination lines of hydrogen (Br\,$\gamma$) and He\,{\sc i}, as well as H$_2$ emission lines. These lines can be produced both in photoionized nebulae and as a consequence of shocks \citep[][]{bla87, dep94, sut17}. Thus, these objects could be either post-AGB stars with shocks or nascent PNe. In any case, they seem more evolved than the sources in Group A, whose spectra indicate cool giant stars. 

\begin{figure}
      \centering
      \includegraphics[width=0.77\hsize]{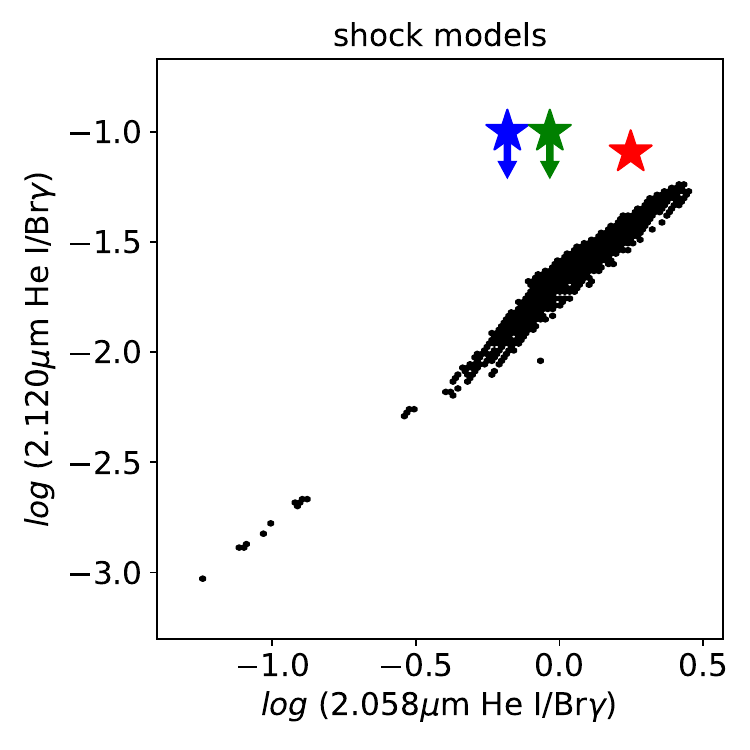}
      \caption{Emission-line ratio diagnostic diagram for recombination lines of Br$\gamma$ at 2.165$\mu$m and He\,{\sc i} lines at 2.058 and 2.120$\mu$m. The region occupied by the black points corresponds to the values expected from the shock models of \cite{sut17}. The filled stars are IRAS\,15103$-$5754 (blue), IRAS\,15445$-$5445 (green), and IRAS\,18443$-$0231 (red). The sizes of the symbols are larger than their errors. The arrows indicate upper limits to the flux of the He\,{\sc i} 2.120\,$\mu$m emission line (see Fig.\,\ref{fig:wf_spectra} and Table\,\ref{tab:lines}).}
      \label{fig:shocks}
\end{figure}

In order to investigate whether the observed emission lines trace photoionized gas typical of PNe or shock-excited gas, we used the Mexican Million Models Database \citep[3MdB;][]{mor15, ala19}, and the shock models by \citet{sut17} stored in the 3MdB \citep[for more details see][]{cal24}. The diagnostic diagram is shown in Fig.\,\ref{fig:shocks} and displays line intensity ratios compatible with shock ionization. Unfortunately, the 3MdB has not yet incorporated the response of the He\,{\sc i} 2.120\,$\mu$m line to photoionization models and the corresponding diagram cannot be constructed in this case.

The diagram in Fig.\,\ref{fig:shocks} does no provide any constraints to the nature of the emission in IRAS\,15103$-$5754 and IRAS\,15445$-$5445 where the He\,{\sc i} at 2.120$\mu$m is not detected in our data. However, we note that the shock-ionization models are distributed along a relatively narrow strip in the diagram, providing a useful prediction of the expected intensity of the He\,{\sc i} 2.120$\mu$m emission line.  Thus, new deeper spectra of these sources could shed light on these objects, by checking whether or not the 2.120$\mu$m\,He\,{\sc i}/Br$\gamma$ ratio is consistent with shocks.

However, in the case of IRAS\,18443$-$0231, this line ratio seems to be incompatible with shocks alone. This strongly suggests that there should be a significant contribution of photoionization in the observed spectrum of this source. Moreover, the He\,\textsc{i} emission at 2.16\,$\mu$m is a factor of $\simeq 2$ stronger than the Br$\gamma$ one, which would be consistent with a photoionized region with  $\log T_{\rm eff} {\rm [K]}$\,$\leq$\,4.6 and $\log n_{e} {\rm [cm^{-3}]}$\,$\geq$\,4, according to Figure 2 of \cite{dep94}. In summary, the  2.120\,$\mu$m\,He\,{\sc i}/Br$\gamma$ ratio suggests that IRAS\,18443$-$0231 could be a nascent PN, in which its radio continuum emission is dominated by non-thermal processes, as in IRAS\,15103$-$5754 \citep[][]{sua15}.

\subsection{The slow and fast jets in WFs}

WFs are usually identified in single-dish spectra of H$_{2}$O maser emission, when they show components spread over a velocity range larger than that expected from the expansion of the AGB envelope \citep[$\simeq$\,50\,km\,s$^{-1}$, assuming maximum expansion velocity $\simeq$\,25\,km\,s$^{-1}$; e.g.,][]{tel91}. However, another criterion commonly used for WF identification is that the H$_{2}$O maser components fall outside the velocity range covered by OH masers \citep{yung13, fan24}, since this is considered to be an indication of the presence of non-spherical mass-loss. This is based on the fact that, in the case of maser emission in spherical AGB envelopes, H$_{2}$O masers are pumped at distances closer to the star than the OH ones. Therefore, their velocities must be lower, with OH masers tracing the terminal velocities of the outer envelope. Similarly, the comparison of the velocity ranges of other lines, such as the OH main lines (1665 and 1667 MHz) with respect to the satellite one (1612 MHz) have been used to infer the presence of non spherical mass-loss \citep[][]{xie25}. In the case of WFs, sources with H$_2$O maser emission that have low velocity spreads ($< 50$\,km\,s$^{-1}$), but fall outside the range of OH velocities, have been classified as `low-velocity' WFs \citep{yung13}. Moreover, independently of the characteristics of OH emission, one could also classify as `low-velocity' WFs those sources whose H$_2$O maser emission shows low velocity spreads, but it is seen to trace bipolar jets, when observed with radio interferometers \citep[e.g.][]{bob05}. These slow jets traced by H$_2$O masers have been interpreted in some cases as a projection effect, when the jet axis is close to the plane of the sky \citep[][]{bob05, vle14}. However, \cite{yung13} found four candidate `low-velocity' WFs, whose infrared colors resemble those of AGB stars, clearly different from classical `high-velocity' WFs, whose colors are typical of post-AGB stars. Based on that, \cite{yung13} suggested that these `low-velocity' WFs are transitional objects in the late AGB or early post-AGB phase, and that their jets are intrinsically slower than those of high-velocity WFs.

\citet{cal25} presented maps of the H$_2$O maser distribution of the three evolved stars of Group A in this paper: IRAS\,18019$-$2216, IRAS\,18464$-$0140, and IRAS\,18480+0008. The first two objects clearly present a linear distribution of the H$_{2}$O maser emission, indicating the presence of jets with radial velocity spreads of only $\leq$30 km\,s$^{-1}$. Thus, they classify them as `low-velocity' WFs. The H$_2$O maser distribution in IRAS\,18480+0008 also suggests a jet, but higher-resolution observations are necessary for confirmation, so we can consider it a candidate `low-velocity' WF. However, the sources in Group B are classical `high-velocity' WF, since the reported velocity spreads of the H$_2$O maser emission in Group B IRAS\,15103$-$5754, IRAS\,15445$-$5449, and IRAS 18443$-$0231 has been reported to be 70, 90, and 80\,km\,s$^{-1}$, respectively \citep[][]{ps13, gom15, cal25}.

The clear differences in the infrared spectra of the Group A sources with respect to the classical `high-velocity' WFs of Group B (Fig.\,\ref{fig:wf_spectra}) strongly suggest that the difference between low- and high-velocity WFs is not necessarily due to a projection effect (although it could be possible in particular cases), but they are intrinsically different objects, in different evolutionary phases. This aligns with the suggestion put forward by \citet[][see above]{yung13}, but we provide a stronger support based on spectroscopy. Moreover, our results may indicate a possible evolution of the ejection velocity of jets in evolved stars, which increases as the post-AGB evolution proceeds. 

We note, however, that there is a potential bias in our sample, since all three sources in our Group B present non-thermal radio continuum emission and evidence for the presence of ionized gas (due to the detection of infrared recombination lines). These properties might not be shared by all known 'high-velocity' WFs. Extending our spectroscopic study to a larger sample of WFs would help us to confirm the evolutionary trends suggested in this work.

\subsection{Origin of the CO and Na\,{\sc i} emission in IRAS\,15445$-$5449} 
\label{sec:co_em}

IRAS\,15445$-$5449 is the only source in our sample that shows the CO band and Na\,{\sc i} doublet in emission. CO band emission has been found in several post-AGB stars, which has been interpreted in some cases as tracing ongoing mass-loss \citep[e.g.,][]{hri94}. In the WF IRAS 16342$-$3814 \citep{cla09}, the CO band and Na\,{\sc i} doublet are both observed in emission \citep{gle12}, as in IRAS\,15445$-$5449. These features originate from an unresolved region around the central star of IRAS 16342$-$3814 and seem to be scattered toward our line of sight by dust in the nebula. \cite{gle12} suggested that this emission may also arise from a circumstellar disk, in analogy to what has been observed in young stellar objects \citep[e.g.,][]{mar97, fed20}. The maser jet in IRAS\,15445$-$5449 \citep[][]{ps13} suggests the presence of both a circumstellar accretion disk and active mass loss, which could be the origin of the CO band and Na\,{\sc i} doublet. Further high-resolution observations can help unravel the spatial location and nature of this emission in IRAS\,15445$-$5449.

\section{Conclusions}

We have presented near-infrared spectra of six objects with WF characteristics that are also emitters of radio continuum, aiming at unveiling their evolutionary stage. The main conclusions of this paper are:

\begin{itemize}

\item The objects IRAS\,18019$-$2216, IRAS\,18464$-$0140, and IRAS\,18480+0008 exhibit CO absorption bands characteristic of cool ($T_{\rm eff}$$\simeq$2100-5700\,K) giant stars, while IRAS 15103$-$5754, IRAS\,15445$-$5449, and IRAS\,18443$-$0231 show H$_2$, Br$\gamma$, and He\,{\sc i} emission lines consistent with shocks around post-AGB stars or photoionization from PNe. In particular, the emission line ratios in IRAS\,18443$-$0231 indicate that it may have entered the photoionization phase and be a nascent PN.

\item The three WFs classified as giant stars are `low-velocity' WFs, while the other three are classical, `high-velocity' WFs, which strongly suggests that the difference between low- and high- velocity WFs reflects intrinsically different evolutionary stages. Moreover, the combined spectral characteristics and maser kinematics support the idea of an evolution of the ejection velocity of jets in evolved stars, with faster ones as the evolution along
the post-AGB phase proceeds.

\item The spectrum of IRAS\,15445$-$5449 shows the CO band in emission and Na\,{\sc i} emission lines, suggesting the presence of a circumstellar disk and/or active mass loss. 
\end{itemize}

\begin{acknowledgements}
Based on observations collected at the European Organisation for Astronomical Research in the Southern Hemisphere under ESO programme(s) 111.24T1.001. This publication makes use of data products from the 2MASS, which is a joint project of the University of Massachusetts and the Infrared Processing and Analysis Center/California Institute of Technology, funded by the National Aeronautics and Space Administration and the National Science Foundation, and UKIDSS. RAC, LFM, JFG are supported by grants PID2023-146295NB-I00 and CEX2021-001131-S, funded by MCIN/AEI /10.13039/501100011033. RAC also acknowledges support by the predoctoral grant PRE2018-085518, funded by MCIN/AEI/ 10.13039/501100011033 and by ESF Investing in your Future. K.O. acknowledges the support of the Agencia Nacional de Investigación Científica y Desarrollo (ANID) through the FONDECYT Regular grant 1240301.

\end{acknowledgements}

\begin{appendix}

\section{Parameters of the observations}
\label{ap:obs}

The \textit{K}-broadband magnitudes of the targets were taken from the Two Micron All Sky Survey (2MASS) and the United Kingdom Infrared Deep Sky Survey (UKIDSS \footnote{The project is defined in \cite{law07} and uses the UKIRT Wide Field Camera \citep[][]{cas07} and a photometric system described in \cite{hew06}. The pipeline processing and science archive are described in \cite{ham08})}). The targets and parameters of the performed observations are shown in Table\,\ref{tab:ntt_obs}.

The small angular size of the targets allowed the creation of sky and target frames simultaneously by jittering (20$\arcsec$ randomly) and nodding (80$\arcsec$ along the slit) the NTT around the target position, without the need of further offset observations of the sky. 

In order to favor the detection of weak emission lines, the Hawaii array was configured in non-destructive mode to take 30 readings per exposure and 4 samples per reading. The detector integration time (DIT) and the number of frames (NDIT) per jittered position were chosen depending on the broadband \textit{K} magnitude of the target, night-time conditions, and field crowding. All targets were observed at least 2 NDIT per jittered position to enhance the signal-to-noise ratio of the emission.

\begin{table*}\setlength{\tabcolsep}{3.5pt}
          \caption[]{Parameters of the near-infrared \textit{K}-band spectroscopic observations.}
          \label{tab:ntt_obs}
          \centering                          
    \begin{tabular}{lccccccccccc}     
    \hline\hline                 
     & \textit{K} $^a$ & \multicolumn{2}{c}{Coordinates$^b$} &  & & Time$^{d}$ & PA$^e$ & \\
     Target  name & (mag) & RA(J2000) & DEC(J2000) & DIT$^c$ & NDIT$^c$ & (min) & ($\degr$) & UTC Date\\
    \hline

    IRAS 15103$-$5754 & 10.64  & 15:14:18.54 & $-$58:05:20.38 &  360 & 2 & 48 & +56 & 11/06/2023 \\
    IRAS 15445$-$5449 & 10.89  & 15:48:19.42 & $-$54:58:20.15 &  360 & 2 & 48 & 0 & 11/06/2023  \\
    IRAS 18019$-$2216 & 14.35  & 18:34:57.36 & $-$22:15:50.84 &  300 & 3 & 60 & 0 & 12/06/2023\\
    IRAS 18443$-$0231 & 13.33  & 18:47:00.40 & $-$02:27:52.47 & 360 & 2 & 48 & 0 & 11/06/2023  \\
    IRAS 18464$-$0140 &  16.24  & 18:48:56.52 & $-$01:36:43.32 & 800 & 2 & 106 & 0 & 11/06/2023  \\
    IRAS 18480+0008 & 7.12 &  18:50:36.69 & +00:12:28.34 & 30 & 4 & 11 & 0 & 12/06/2023 \\
    \hline
    \end{tabular}
    
    Notes:$^{a}$ Broadband \textit{K} magnitudes were taken from 2MASS, except for that of IRAS\,18019$-$2216, which was taken from UKIDSS. $^b$ Coordinates of the radio continuum emission reported in \citet{sua15} and \citet{cal22, cal25}. $^{c}$ Detector integration time in seconds (DIT), and number of frames (NDIT). $^d$ Total integration time. $^e$ Position angle of the slit.
    
        \end{table*}

\section{Emission line fluxes of Group B sources} 
\label{ap:lines}

In Table\,\ref{tab:lines} we report the emission line fluxes in the spectra of Group B: IRAS\,15103$-$5754, IRAS\,15445$-$5445, and IRAS\,18443$-$0231 (Fig.\,\ref{fig:wf_spectra}). These values have been obtained by fitting gaussians following standard procedures in the software IRAF\footnote{IRAF is distributed by the National Optical Astronomy Observatories, which is operated the Association of Universities for Research in Astronomy, Inc., under contract to the National Science Foundation.}.

\begin{table}[b]
\caption{Emission line fluxes F($\lambda$) in units of F(Br$\gamma$)=100 in the spectra of Group B sources.} \label{tab:lines}
\centering
\begin{tabular}{lccc}
\hline \hline
    & IRAS15103\tablefootmark{a}  & IRAS15445\tablefootmark{b}  & IRAS18443\tablefootmark{c}  \\
Emission line   &  F($\lambda$)\tablefootmark{d}  &  F($\lambda$)\tablefootmark{d}  & F($\lambda$)\tablefootmark{d} \\

\hline
2.025	H$_2$      	&	--			&	12.9$\pm$2.3	&	--			\\
2.035	H$_2$      	&	89.0$\pm$5.5	&	32.6$\pm$3.6	&	5.7$\pm$1.2		\\
2.046	H$_2$      	&	38.9$\pm$3.5	&	46.3$\pm$4.4	&	--			\\
2.060	He\,{\sc i}	&	65.8$\pm$4.9	&	92.5$\pm$6.6	&	176.9$\pm$6.9			\\
2.075	H$_2$      	&	43.7$\pm$3.5	&	17.0$\pm$2.6	&	--			\\
2.120	He\,{\sc i}	&	--			   &	--	             &	8.1$\pm$2.2			\\
2.122	H$_2$      	&	264.5$\pm$8.9	&	117.0$\pm$7.2	&	19.9$\pm$2.3			\\
2.164	Br$\gamma$	&	100$\pm$7	&	100$\pm$9	&	100$\pm$3		\\
2.210 Na I       &       --        &    26.4$\pm$3.1   &     -- \\
2.221	H$_2$	&	150.1$\pm$6.4  	&	137.5$\pm$7.6	&	2.73$\pm$0.29			\\
2.240    H$_2$      &       --        &    17.1$\pm$2.6  &       --    \\
2.246	H$_2$	&	25.3$\pm$2.7	&	22.4$\pm$3.0	& 	-- \\
2.289	CO	&	--	&	89.5$\pm$5.9	&	--			\\
\hline
F(Br$\gamma$)$^e$ & 4.80$\pm$0.32 & 2.00$\pm$0.19 & 1.730$\pm$0.044\\

\hline 
\hline                                   
\end{tabular}
\tablefoot{ \tablefootmark{a}{IRAS\,15103$-$5754.}
            \tablefootmark{b}{IRAS\,15445$-$4559.}
            \tablefootmark{c}{IRAS\,18443$-$0231.}
            \tablefootmark{d}{'--' stands for not detected.}
            \tablefootmark{e}{In units of 10$^{-18}$\,W\,m$^{-2}$.}
                        }	
		
\end{table}

\end{appendix}
\end{document}